\documentclass
[english,aps,floats,twocolumn,showpacs,nofootinbib,floatfix]
{revtex4}
\usepackage{pslatex}
\usepackage[T1]{fontenc}
\usepackage[latin1]{inputenc}
\usepackage{graphicx}
\usepackage{epsfig}
\usepackage{longtable}
\usepackage{float}

{
{
{
\newcommand{\bea}{\begin{eqnarray}}
\newcommand{\eea}{\end{eqnarray}}

\newcommand{\nc}{\newcommand}
\nc{\renc}{\renewcommand}
\nc{\eqs}[2]{\mbox{Eqs.~(\ref{#1},\,\ref{#2})}}
\nc{\eq}[1]{\mbox{Eq.~(\ref{#1})}}
\nc{\figs}[2]{\mbox{Figs.~(\ref{#1},\,\ref{#2})}}
\nc{\fig}[1]{\mbox{Fig~.(\ref{#1})}}
\nc{\be}[1]{\begin{equation} \mbox{$\label{#1}$}}
\nc{\ee}{\vspace{0.1cm}\end{equation}}

\newcommand{\bean}{\begin{eqnarray*}}
\newcommand{\eean}{\end{eqnarray*}}

\def\bfn{{\bf n}}
\def\bfp{{\bf p}}

\def\cP{{\cal P}}

\begin{document}
\title{Hemispherical Power Asymmetry from a Space-Dependent Component of the Adiabatic Power Spectrum}
\author{John McDonald}
\email{j.mcdonald@lancaster.ac.uk}
\affiliation{Dept. of Physics, University of 
Lancaster, Lancaster LA1 4YB, UK}

\begin{abstract}

    The hemispherical power asymmetry observed by Planck and WMAP can be interpreted as due to a spatially-varying and scale-dependent component of the adiabatic power spectrum. We derive general constraints on the magnitude and scale-dependence of a component with a dipole spatial variation.  The spectral index and the running of the spectral index can be significantly shifted from their inflation model values, resulting in a smaller spectral index and a more positive running. A key prediction is a hemispherical asymmetry of the spectral index and of its running. Measurement of these asymmetries can test the structure of the perturbation responsible for the CMB power asymmetry.

\end{abstract}
 \maketitle

\section{Introduction} 

      The Planck satellite has observed a hemispherical asymmetry in the CMB temperature fluctuations at low multipoles \cite{planckasym}, confirming the earlier observation by WMAP \cite{wmapearly,wmap5}.
 The asymmetry can be modelled by a temperature fluctuation dipole of the form \cite{gordon}
\be{e1} \frac{\delta T}{T}(\hat{\bfn}) = \left(\frac{\delta T}{T}\right)_{o}(\hat{\bfn})\left[1 + A\;\hat{\bfn}.\hat{\bfp} \right]   ~,\ee  
where $\left(\frac{\delta T}{T}\right)_{o}(\hat{\bfn})$ is a statistically isotropic temperature fluctuation, $A$ is the magnitude of the asymmetry and $\hat{\bfp}$ is its direction. Recent Planck results give $A = 0.073 \pm 0.010$ in the direction $(217.5 \pm 15.4, -20.2 \pm 15.1)$ for multipoles $l \in (2, 64)$ \cite{planckasym}.  This asymmetry is unlikely to arise as a result of random  fluctuations in a statistically isotropic model, with less than one out of a thousand isotropic simulations fitting the asymmetry observed by Planck \cite{hansennew}. Analyses and proposed explanations of the hemispherical power asymmetry are discussed in \cite{kam1,kam2,kam3,pesky,lyth,jm1,jm2,list}.

    An important constraint on such models is the absence of an asymmetry at smaller angular scales. In particular, the asymmetry on scales corresponding to quasar number counts must satisfy $A < 0.012$ at 95$\%$ c.l. \cite{quasar}, while a more recent analysis of Planck data finds that $A < 0.0045$ at 95$\%$ c.l. for $l = 601-2048$ \cite{sh}. 

    A natural interpretation of these observations, which we discuss in this paper, is the existence of an additional space-dependent adiabatic component of the curvature power spectrum. This must be strongly scale-dependent in order to suppress the asymmetry on small angular scales. We will consider in the following the case of an additional adiabatic component with a dipole spatial variation.

\section{Hemispherical Asymmetry from a Dipole Component of the Adiabatic Power Spectrum}

  We will consider a component of the adiabatic power spectrum whose magnitude is a function of angle $\theta$ on the surface of last scattering,   
\be{e2} \cP_{\zeta} = \cP_{inf} + \cP_{asy}  ~.\ee
Here $\cP_{\zeta}$ is the power spectrum extracted from a region around a point at angle $\theta$ on the last-scattering surface 
\cite{wmapearly,wmap5}, $\cP_{inf}$ is the conventional inflaton power spectrum and $\cP_{asy}$ is the additional scale-dependent adiabatic component responsible for the hemispherical asymmetry. $\cP_{asy}$ consists of a mean value $\hat{\cP}_{asy}$ and a spatial variation about this mean of magnitude $\Delta \cP_{asy}$,
\be{e3}  \cP_{asy} = \hat{\cP}_{asy}  +  \Delta \cP_{asy} \cos \theta ~,\ee
where $\cos \theta = \hat{\bfn}.\hat{\bfp}$.
This corresponds to an adiabatic power spectrum component 
with a dipole term in the direction $\hat{\bfp}$. 

   To relate the asymmetry $A$ to the curvature power spectrum, we will compute the mean squared temperature fluctuation as a function of $\theta$. This is determined by the curvature power on the last-scattering surface at $\theta$, which can be related to the corresponding multipoles via $C_{l}(\theta) = \cP_{\zeta}(k,     
 \theta) \hat{C}_{l}$, where $\hat{C}_{l}$ is the adiabatic perturbation multipole for a scale-invariant spectrum with   
 $\cP_{\zeta} = 1$ \cite{kurki} and $C_{l}(\theta)$ are the  modulated multipoles as a function of $\theta$. Each multipole $C_{l}$ receives contributions from a range of $k$ around $k = l/x_{ls}$, where $x_{ls} = 14100$ Mpc is the comoving distance to the last-scattering surface. The range of $k$ is sufficiently narrow that the effect of the scale-dependence of the power spectrum can be accurately estimated by setting  $k$ to $l/x_{ls}$ in 
 $\cP_{\zeta}(k, \theta)$. We define $\overline{C}_{l}$ to correspond to $\theta = \pi/2$ and $\Delta C_{l}(\theta)$ to be the change as a function of $\theta$. Then, for multipoles in the range $l_{min}$ to $l_{max}$, we obtain  
\be{e7} \frac{\Delta \left( \frac{\delta T}{T}  \right)_{\theta}^{2}}{\left( \frac{\delta T}{T}  \right)_{o}^{2}} =  
\frac{ \displaystyle\sum_{l=l_{min}}^{l_{max}} \left(2 l + 1 \right) \Delta C_{l}(\theta) }{\displaystyle\sum_{l=l_{min}}^{l_{max}} \left(2 l + 1 \right) \overline{C}_{l}   }    ~.\ee
   In practice, a binned power spectrum, which we will denote by $\tilde{C}_{l}$, is extracted from the temperature data, where $l(l+1)\tilde{C}_{l}$ is a constant for each bin \cite{hivon,hansen02}. We therefore need to estimate $\tilde{C}_{l}$ 
from the true $C_{l}$ for a given perturbation. 
To do this we match 
the mean squared temperature fluctuation calculated with $C_{l}$ to that calculated with $\tilde{C}_{l}$. In this case $\sum (2l+1) \tilde{C}_{l} 
= \sum (2l+1) C_{l}$ for each bin. $\tilde{C}_{l}$ for the bin $l = l_{min}$ to $l_{max}$ is therefore given by
\be{cl1}   \tilde{C}_{l} = \frac{1}{l(l+1)} \times \frac{\displaystyle\sum_{l^{\prime} = l_{min}}^{l_{max}} 
(2l^{\prime}+1)  C_{l^{\prime}}}{\displaystyle\sum_{l^{\prime} = l_{min}}^{l_{max}} \frac{(2 l^{\prime} + 1)}{l^{'}(l^{'} + 1)} }  ~.\ee
The observed asymmetry $A$ in a given bin is derived from the asymmetry in the corresponding $\tilde{C}_{l}$. We will therefore replace $C_{l}$ by $\tilde{C}_{l}$ in \eq{e7}. 
   To obtain $A$ we compare \eq{e7} with the value expected from the temperature fluctuation dipole \eq{e1},
\be{e6} \frac{\Delta \left( \frac{\delta T}{T}  \right)_{\theta}^{2}}{\left( \frac{\delta T}{T}  \right)_{o}^{2}} \approx 2 (\hat{\bfn}.\hat{\bfp})A  ~.\ee 
where we assume that $A \ll 1$.  We define $\overline{\cP}_{\zeta} = \cP_{inf} + \hat{\cP}_{asy}$ to be the adiabatic power at $\theta = \pi/2$.  Then $\Delta C_{l}(\theta)/\overline{C}_{l} = (\hat{\bfn}.\hat{\bfp}) \Delta \cP_{asy}(k)/\overline{\cP}_{\zeta}(k)$, with $k$ corresponding to $l$. By comparing \eq{e7} (with $C_{l} \rightarrow \tilde{C}_{l}$) and \eq{e6}, we obtain
 \be{e8} A = \frac{ 
\displaystyle \sum_{l = l_{min}}^{l_{max}} \frac{(2l+1)}{l(l+1)} 
\displaystyle \sum_{l^{\prime} = l_{min}}^{l_{max}} (2 l^{\prime} + 1)   \left( \frac{\Delta \cP_{asy}(k^{\prime})}{\overline{\cP}_{\zeta}(k^{\prime})} \right)    \overline{C}_{l^{\prime}} }{
2 \displaystyle \sum_{l = l_{min}}^{l_{max}} \frac{(2l+1)}{l(l+1)}  \displaystyle \sum_{l^{\prime} = l_{min}}^{l_{max}} (2 l^{\prime} + 1) \overline{C}_{l^{\prime}}    }     ~, \ee
where $k^{\prime} = l^{\prime}/x_{ls}$.  
In the following we will assume that $\xi \ll 1$, where $\xi = \hat{\cP}_{asy}/\cP_{inf}$, and work to leading order in $\xi$. Then 
\be{e9} \frac{ \Delta \cP_{asy}}{\overline{\cP}_{\zeta}} = 
\frac{\xi}{ \left(1 + \xi\right) }  \frac{ \Delta \cP_{asy}}{\hat{\cP}_{asy}} \approx \xi  \frac{ \Delta \cP_{asy}}{\hat{\cP}_{asy}} ~.\ee 
     In general, the scale-dependence of $\hat{\cP}_{asy}$ may be different from the scale-dependence of the spatial change of the power $\Delta \cP_{asy}$. We will therefore introduce different spectral indices to parameterize these\footnote{In this study we will assume that the spectral indices $n_{\sigma}$ and $n_{\Delta}$ are not significantly running. Generalizations of $\cP_{\zeta}$ will be considered in future work.},
\be{e10} \hat{\cP}_{asy} = \hat{\cP}_{asy\;\;0} \left(\frac{k}{k_{0}}\right)^{n_{\sigma} - 1}
\;\;\;;\;\;\;  
\frac{\Delta \cP_{asy}}{\hat{\cP}_{asy} } = \left( \frac{\Delta \cP_{asy}}{\hat{\cP}_{asy} } \right)_{0}
\left(\frac{k}{k_{0}}\right)^{n_{\Delta} - 1}    ~,\ee
where subscript $0$ denotes values at the pivot scale $k_{0}$. 
If the space-dependence of the curvature power $\Delta \cP_{asy}$ has the same scale-dependence as $\hat{\cP}_{asy}$ then $n_{\Delta}  = 1$ \footnote{This is true, for example,  for the modulated reheating model of \cite{jm2}, which also predicts no running of $n_{\sigma}$.}.  

In the following we will use the Planck pivot scale, $k_{0} = 0.05 {\rm Mpc}^{-1}$. In this case the corresponding multipole number is $l_{0} = 700$. Setting $(k/k_{0}) =  (l/l_{0})$ in \eq{e10} then gives a good estimate of the scale-dependence.  We will assume that the scale-dependence of the inflaton perturbation is negligible compared to that of $\hat{\cP}_{asy}$. \eq{e8} then becomes, 
\be{e11} A = \frac{\xi_{0} 
(\Delta \cP_{asy}/\hat{\cP}_{asy})_{0} 
}{2} \times 
\frac{ \displaystyle \sum_{l = l_{min}}^{l_{max}} (2 l + 1)   
\left(\frac{l}{l_{0}}\right)^{n_{A} - 2} 
\overline{C}_{l} }{
\displaystyle \sum_{l = l_{min}}^{l_{max}} (2 l + 1)\overline{C}_{l}    }     ~,\ee
where $n_{A} = n_{\sigma} + n_{\Delta}$. 
In this we are assuming that $\overline{C}_{l}$ is dominated by the inflaton perturbation, which can be considered to be scale-invariant here. 

  For $l$ from 2 to $l_{max} = 64$, $l(l+1)\overline{C}_{l}$ has only a small variation. We can therefore consider $l(l+1)\overline{C}_{l}$ to be approximately constant, in which case the large-angle asymmetry observed by Planck and WMAP, which we will denote by $A_{large}$, is  given by
\be{e12} A_{large} \approx \frac{\xi_{0} (\Delta \cP_{asy}/\hat{\cP}_{asy})_{0}}{2} \times \frac{ \displaystyle\sum_{l=2}^{64} \frac{\left(2 l + 1 \right)}{l\left(l+1\right)} \left(\frac{l}{l_{0}}\right)^{n_{A} - 2} }{\displaystyle\sum_{l=2}^{64} \frac{\left(2 l + 1 \right)}{l\left(l+1\right)}  }  ~.\ee
A recent analysis of Planck data finds that on smaller angular scales the asymmetry satisfies $A < 0.0045$ (95$\%$ c.l.) for $l = 601-2048$ \cite{sh}. This is stronger than the earlier quasar bound, $A <  0.012$ (95$\%$ c.l.) on scales $k = (1.3 - 1.8) h \; {\rm Mpc^{-1}}$, corresponding to  $l = 12400 - 17200$ \cite{quasar}, and is consistent with an analysis of the trispectrum from Planck, which finds $A \sim 0.002$ at $l \approx 2000$ \cite{PlanckNG}. For large $l$ we can integrate the sums in \eq{e11} over $l$. In this case 
\be{e14} A \approx 
\frac{\xi_{0} (\Delta \cP_{asy}/\hat{\cP}_{asy})_{0}}{2} 
\times 
\frac{ \left(\left(\frac{l_{max}}{l_{min}}\right)^{n_{A} - 2} - 1 \right)}{\left(n_{A} - 2\right) \ln \left(\frac{l_{max}}{l_{min}}\right) } 
\left(\frac{l_{min}}{l_{0}}\right)^{n_{A} - 2}   ~,\ee
We then define the small-angle asymmetry, $A_{small}$,  to be given by
\eq{e14} with $l_{min} = 601$ and $l_{max} = 2048$.

\section{The Spectral Index and its Running}

       A general consequence of an additional scale-dependent adiabatic component of the power spectrum is that the spectral index and the running of the spectral index will be modified from their inflation model values. The power spectrum and spectral index are determined by the mean-squared CMB temperature fluctuations over the whole sky.  
This can be thought of as the average of the mean-squared temperature fluctuations at different $\theta$. Since from \eq{e6} the mean-squared temperature fluctuation at $\pi/2 + \Delta \theta$ cancels that from $\pi/2-\Delta \theta$, the mean power from averaging over all angles $\theta$ will be equal to the power at $\theta = \pi/2$, 
\be{e17} \overline{\cP}_{\zeta} =   \cP_{inf} + \hat{\cP}_{asy}  ~.\ee
The spectral index as observed by Planck, $n_{s}$,  is therefore given by 
\be{e19} n_{s} - 1  =  \frac{k}{\overline{\cP}_{\zeta}} \frac{d \overline{\cP}_{\zeta}}{dk} 
= \frac{\left(n_{s} - 1\right)_{inf}}{\left(1 + \xi\right)}
   + \frac{\xi}{\left(1 + \xi\right)} \left(n_{\sigma} - 1\right) ~,\ee
where $n_{s\;inf} = (k/\cP_{inf})(d\cP_{inf}/dk)$. The running of the spectral index, $n_{s}^{\prime}$, is given by 
\be{e19a} n_{s}^{\prime} \equiv \frac{d n_{s}}{d \ln k}  
=  \frac{n_{s\; inf}^{\prime}}{\left( 1 + \xi\right)} 
+ \frac{\xi}{\left(1+\xi\right)^{2}} \left( n_{\sigma} - n_{s\; inf} \right)^{2}   ~.\ee
To leading order in $\xi$ we therefore find that 
$n_{s} - 1 \approx \left(n_{s} - 1\right)_{inf} + \Delta n_{s}$
and $n_{s}^{\prime} \approx n_{s\;inf}^{\prime} +  \Delta n_{s}^{\prime}$,  where 
\be{e21} \Delta n_{s} = \xi \left( \left(n_{\sigma} - 1\right) - 
\left(n_{s} - 1\right)_{inf} \right)  ~\ee
and
\be{e22} \Delta n_{s}^{\prime} = 
\xi \left(\left(n_{\sigma} - n_{s\;inf}\right)^{2} - n_{s
\;inf}^{\prime}  
\right)   ~.\ee

\section{Hemispherical Asymmetry of the Spectral Index and its Running}

   There is also a hemispherical asymmetry in the spectral index and the running of the spectral index, obtained by averaging the temperature fluctuations over each hemisphere. For the hemisphere from $\theta = 0$ to $\theta = \pi/2$, which we denote by $+$, the 
average power is   
\be{e23} \overline{\cP}_{\zeta\;+}
\equiv \int_{0}^{\pi/2} \left(\cP_{inf} + \hat{\cP}_{asy} + \Delta \cP_{asy} \cos \theta \right) \sin \theta d \theta  
~.\ee  
Therefore
\be{e23a} \overline{\cP}_{\zeta\;+}
= \overline{\cP}_{\zeta} + \frac{1}{2} \Delta P_{asy} ~.\ee
For the opposite hemisphere, $\overline{\cP}_{\zeta\;-} = \overline{\cP}_{\zeta} - \frac{1}{2} \Delta P_{asy} $. 
The spectral index from the average power in each hemisphere, $n_{s\;\pm}$, is therefore 
\be{e24} n_{s\;\pm} - 1 = \frac{k}{\overline{\cP}_{\zeta\;\pm}} \frac{d  \overline{\cP}_{\zeta\;\pm}}{dk}   ~.\ee
Assuming that $\Delta \cP_{asy}/2 \ll \overline{\cP}_{\zeta}$ and neglecting the scale-dependence of $\cP_{inf}$, we find  
that $n_{s\;\pm}  \approx n_{s} \pm \delta n_{s} $ where
\be{e25} \delta n_{s} = \frac{\xi_{0} (\Delta \cP_{asy}/\hat{\cP}_{asy})_{0}}{2} \left(n_{A} - 2\right) \left(\frac{k}{k_{0}}\right)^{n_{A} - 2}  ~.\ee 
Similarly, for the running of the spectral index we find that
$ n_{s\;\pm}^{\prime} \approx n_{s}^{\prime} \pm \delta n_{s}^{\prime} $, where
\be{e27} \delta n_{s}^{\prime}  = \frac{\xi_{0} (\Delta \cP_{asy}/\hat{\cP}_{asy})_{0}}{2} \left(n_{A} - 2\right)^{2} \left(\frac{k}{k_{0}}\right)^{n_{A} - 2}  ~.\ee 

The spectral index parameters for the power spectrum over a hemisphere can be extracted from the CMB data in much the same way that Planck determines the parameters for the whole sky. Therefore a similar level of accuracy can be expected. These parameters will completely characterize the CMB fluctuations over a hemisphere in a model-independent way. The spectral index parameters over a hemisphere can then be used to test specific models for the power asymmetry, such as \eq{e3} combined with \eq{e10}, by comparing with their predicted values.

\section{Results} 

    In Table 1 we give the values of $A_{small}$ and $\xi_{0} (\Delta \cP_{asy}/\hat{\cP}_{asy})_{0}$ as a function of $n_{A} = n_{\sigma} + n_{\Delta}$, where have fixed $A_{large}$ to its observed value 0.073 throughout. We find that $n_{A} < 1.44$ is necessary to have a strong enough scale-dependence to satisfy the Planck bound $A_{small} < 0.0045$. $\xi_{0} (\Delta \cP_{asy}/\hat{\cP}_{asy})_{0}$ decreases with $n_{A}$ from a maximum value of $0.012$ at $n_{A} = 1.44$.

\begin{table*}
\begin{center}
\begin{tabular}{|c|c|c|c|c|c|c|}
 \hline $n_{A}$	 &  2.0 & 1.5 & 1.44 & 1.2 & 1.0 & 0.5 \\
\hline	$A_{small}$	&	$0.073$ & $0.0063$ & $0.0045$ & $0.0013$ & $4.4 \times 10^{-4}$ & $2.7 \times 10^{-5}$  \\
\hline	$\xi_{0} (\Delta \cP_{asy}/\hat{\cP}_{asy})_{0}$	&	$0.146$ & $0.016$ & $0.012$ & $0.0036$ & $0.0013$  & $9.3 \times 10^{-5}$ \\
\hline	$\Delta n_{s}/\xi$	&	$0.0$ & $-0.50$  & $-0.56$  & $-0.80$ & $-1.00$ & $-1.50$  \\
\hline	$\Delta n_{s}^{\prime}/\xi$	&	$0.0$	& $0.25$ & $0.31$ & $0.64$ & $1.00$ & $2.25$ \\
\hline	$\delta n_{s} (l = 28)$	&	$0.0$	 & $-0.019$ & $-0.020$  & $-0.019$ & $-0.016$ &  $-0.0087$\\
\hline	$\delta n_{s}^{\prime} (l = 28)$	&	$0.0$	 &  $0.0097$ & $0.011$ & $0.015$ & $0.016$  & $0.013$\\
\hline	$\delta n_{s} (l = 700)$	&	$0.0$	 & $-0.0039$ & $-0.0033$  & $-0.0014$ & $-6.5 \times 
10^{-4}$ &  $-7.0 \times 10^{-5}$\\
\hline	$\delta n_{s}^{\prime} (l = 700)$	&	$0.0$	 &  $0.0019$ & $0.0018$ & $0.0012$ & $6.5 \times 
10^{-4}$  & $1.0 \times 10^{-4}$\\
\hline     
 \end{tabular} 
 \caption{\footnotesize{$\xi_{0} (\Delta \cP_{asy}/\hat{\cP}_{asy})_{0}$, $A_{small}$ and spectral index parameters at as a function of $n_{A}$. }  }
 \end{center}
 \end{table*}

   We next consider the shift of the spectral index and the running of the spectral index from their inflation model values.  We will consider the case where the scale-dependence is mostly due to $\hat{\cP}_{asy}$ rather than $\Delta \hat{\cP}_{asy}/\hat{\cP}_{asy}$ and therefore set $n_{\Delta} = 1$, in which case $n_{A} = n_{\sigma} + 1$. This gives the maximum shift of the spectral index and its running for a given value of $n_{A}$ and $\xi_{0}$. We also set $n_{s\;inf} = 1$ throughout. Table 1 gives the values of $\Delta n_{s}/\xi$ and  $\Delta n_{s}^{\prime}/\xi$ as a function of $n_{A}$. The spectral index decreases relative to the inflation model value, while the running of the spectral index increases. The shift of the running of the spectral index imposes a strong constraint on $\xi_{0}$. The Planck result is $n_{s}^{\prime} =  -0.013 \pm 0.018$ (Planck + WP) \cite{planckcosmo}. This imposes the 2-$\sigma$ upper bound $\Delta n_{s}^{\prime} < 0.005$, assuming that the running of the inflation model spectral index is negligible. Comparing this bound with the shift in the spectral index at $k_{0}$ when $n_{A} = 1.44$ implies that $\xi_{0} < 0.016$. Combined with $\xi_{0} (\Delta \cP_{asy}/\hat{\cP}_{asy})_{0} = 0.012$ when $n_{A} = 1.44$ implies that $(\Delta \hat{\cP}_{asy}/\hat{\cP}_{asy})_{0} > 0.75$ is necessary in order to account for the power asymmetry while keeping the running of the spectral index at $k_{0}$ below the Planck 2-$\sigma$ upper limit.

 These constraints can be relaxed if $n_{\sigma}$ is increased for a given $n_{A}$ by reducing $n_{\Delta}$. This will depend on the specific model responsible for the additional adiabatic component. Alternatively, the positive shift of the running of the spectral index may simply indicate that the underlying inflation model has a negative running of the spectral index.  

  It is also possible to achieve a significant shift of the spectral index relative to its inflation model value. For the case $n_{A} = 1.44$,  $\xi_{0} < 0.016$ 
implies that $\Delta n_{s} > -0.009$ at $k_{0}$. Therefore the inflation model spectral index can be significantly reduced if $n_{A}$ is close to its upper bound from $A_{small}$ and $\xi_{0}$ is close to its upper bound from the running of the spectral index. 

   We finally consider the hemispherical asymmetry of the spectral index and its running. These are completely fixed by $n_{A}$. (The spectral index of $\Delta \cP_{asy}$ is $n_{A} -1$.)
The asymmetry will be largest at small multipoles, where $\cP_{asy}$ is largest. To show the magnitude at diffferent scales, we have calculated the asymmetries at the WMAP pivot scale ($l \approx 28$) and at the Planck pivot scale ($l \approx 700$). From Table 1 we find that $\delta n_{s}(l = 28)$ is in the range $-0.0087$ to $-0.020$  and $\delta n_{s}^{\prime}(l = 28)$ is in the range $0.011$ to $0.016$ for $n_{A}$ varying between 0.5 and 1.44, while
$\delta n_{s}(l = 700)$ is in the range $-7.0 \times 10^{-5}$ to $-0.0033$  and $\delta n_{s}^{\prime}(l = 700)$ is in the range $1.0 \times 10^{-4}$ to $0.018$.

    Measurement of the spectral index parameters for the power spectrum over a hemisphere provides a strategy for the determination of $\cP_{asy}$, by comparing the model-independent measured values of $\delta n_{s}$ and $\delta n_{s}^{\prime}$ with the values predicted by the proposed form for $\cP_{\zeta}$. For example, in the case of $\cP_{\zeta}$ from \eq{e3} combined with \eq{e10}, $n_{A}$ and $\xi_{0}(\Delta \cP_{asy}/\hat{\cP}_{asy})_{0}$ can be fixed by the observed values of $\delta n_{s}$ and $\delta n_{s}^{\prime}$ via \eq{e25}  and \eq{e27}.  
Since $A_{large}$ in \eq{e12} is also determined by $n_{A}$ and $\xi_{0}(\Delta \cP_{asy}/\hat{\cP}_{asy})_{0}$, comparing the predicted value of $A_{large}$ with the observed value will provide a consistency test for $\cP_{asy}$.

   Our analysis is based only on the power law scale-dependence and dipole variation of the additional adiabatic component of the power spectrum. 
These properties must be explained by specific models for the origin of the additional component. Such models will also have to satisfy additional constraints, in particular those from non-Gaussianity and the isotropy of the CMB temperature, which are beyond the model-independent analysis presented here. 

\section*{Acknowledgements}
The author thanks David Lyth for his comments.
The work of JM is partially supported by STFC grant
ST/J000418/1.


\begin{thebibliography}{99}


\bibitem{planckasym} P.~A.~R.~Ade {\it et al.} [Planck Collaboration], 
arXiv:1303.5083 [astro-ph.CO].


\bibitem{wmapearly}
H.~K.~Eriksen, F.~K.~Hansen, A.~J.~Banday, K.~M.~Gorski and P.~B.~Lilje, 
Astrophys.\ J.\ {\bf 605}, 14 (2004)
[Erratum-ibid.\ {\bf 609}, 1198 (2004)]
[astro-ph/0307507];
 F.~K.~Hansen, A.~J.~Banday and K.~M.~Gorski, 
Mon.\ Not.\ Roy.\ Astron.\ Soc.\ {\bf 354}, 641 (2004)
[astro-ph/0404206].
H.~K.~Eriksen, A.~J.~Banday, K.~M.~Gorski, F.~K.~Hansen and P.~B.~Lilje, 
Astrophys.\ J.\ {\bf 660}, L81 (2007)
[astro-ph/0701089]; 
F.~K.~Hansen, A.~J.~Banday, K.~M.~Gorski, H.~K.~Eriksen and P.~B.~Lilje, 
Astrophys.\ J.\ {\bf 704} (2009) 1448
[arXiv:0812.3795 [astro-ph]];


\bibitem{wmap5} J.~Hoftuft, H.~K.~Eriksen, A.~J.~Banday, K.~M.~Gorski, F.~K.~Hansen and P.~B.~Lilje, 
Astrophys.\ J.\ {\bf 699}, 985 (2009)
[arXiv:0903.1229 [astro-ph.CO]].


\bibitem{gordon}  C.~Gordon, W.~Hu, D.~Huterer and T.~M.~Crawford, 
Phys.\ Rev.\ D {\bf 72} (2005) 103002
[astro-ph/0509301].

\bibitem{hansennew} Y.~Akrami, Y.~Fantaye, A.~Shafieloo, H.~K.~Eriksen, F.~K.~Hansen, A.~J.~Banday and K.~M.~Górski, 
arXiv:1402.0870 [astro-ph.CO].



\bibitem{kam1} A.~L.~Erickcek, M.~Kamionkowski and S.~M.~Carroll, 
Phys.\ Rev.\ D {\bf 78}, 123520 (2008)
[arXiv:0806.0377 [astro-ph]].

\bibitem{kam2}  A.~L.~Erickcek, S.~M.~Carroll and M.~Kamionkowski, 
Phys.\ Rev.\ D {\bf 78}, 083012 (2008)
[arXiv:0808.1570 [astro-ph]].

\bibitem{kam3} A.~L.~Erickcek, C.~M.~Hirata and M.~Kamionkowski, 
Phys.\ Rev.\ D {\bf 80}, 083507 (2009)
[arXiv:0907.0705 [astro-ph.CO]].



\bibitem{pesky}  L.~Dai, D.~Jeong, M.~Kamionkowski and J.~Chluba, 
arXiv:1303.6949 [astro-ph.CO].



\bibitem{lyth} D.~H.~Lyth, 
arXiv:1304.1270 [astro-ph.CO].




\bibitem{jm1} J.~McDonald, 
JCAP {\bf 1307} (2013) 043
[arXiv:1305.0525 [astro-ph.CO]].




\bibitem{jm2} J.~McDonald,
  JCAP {\bf 1311} (2013) 041
  [arXiv:1309.1122 [astro-ph.CO]].

 


\bibitem{list}
L.~Wang and A.~Mazumdar, 
Phys.\ Rev.\ D {\bf 88}, 023512 (2013)
[arXiv:1304 .6399 [astro-ph.CO]];
Z.~-G.~Liu, Z.~-K.~Guo and Y.~-S.~Piao,
arXiv:1304.6527 [astro-ph.CO]; 
M.~H.~Namjoo, S.~Baghram and H.~Firouzjahi, 
arXiv:1305.0813 [astro-ph.CO];
A.~R.~Liddle and M.~Cortes,
  arXiv:1306.5698 [astro-ph.CO].
G.~D'Amico, R.~Gobbetti, M.~Kleban and M.~Schillo, 
arXiv:1306.6872 [astro-ph.CO];
A.~A.~Abolhasani, S.~Baghram, H.~Firouzjahi and M.~H.~Namjoo, 
arXiv:1306.6932 [astro-ph.CO];
A.~Mazumdar and L.~Wang, 
arXiv:1306.5736 [astro-ph.CO];
X.~Gao,
  JCAP {\bf 1310} (2013) 039
  [arXiv:1307.2564 [hep-th]];
Y.~-F.~Cai, W.~Zhao and Y.~Zhang, 
arXiv:1307.4090 [astro-ph.CO];
 P.~K.~Rath and P.~Jain, 
JCAP {\bf 1312} (2013) 014
[arXiv:1308.0924 [astro-ph.CO]].
K.~Kohri, C.~-M.~Lin and T.~Matsuda,
  arXiv:1308.5790 [hep-ph];
S.~Kanno, M.~Sasaki and T.~Tanaka,
  arXiv:1309.1350 [astro-ph.CO];
Z.~-G.~Liu, Z.~-K.~Guo and Y.~-S.~Piao,
  arXiv:1311.1599 [astro-ph.CO];
R.~Fernández-Cobos, P.~Vielva, D.~Pietrobon, A.~Balbi, E.~Martínez-González and R.~B.~Barreiro,
  arXiv:1312.0275 [astro-ph.CO];
Z.~Chang and S.~Wang,
  arXiv:1312.6575 [astro-ph.CO];
L.~Wang,
  arXiv:1401.1584 [astro-ph.CO];
P.~K.~Rath and P.~Jain, 
arXiv:1403.2567 [astro-ph.CO].


\bibitem{quasar} C.~M.~Hirata, 
JCAP {\bf 0909}, 011 (2009)
[arXiv:0907.0703 [astro-ph.CO]].


\bibitem{sh} S.~Flender and S.~Hotchkiss, 
JCAP {\bf 1309} (2013) 033
[arXiv:1307.6069 [astro-ph.CO]].


\bibitem{kurki} H.~Kurki-Suonio, V.~Muhonen and J.~Valiviita, 
Phys.\ Rev.\ D {\bf 71} (2005) 063005 [astro-ph/0412439].



\bibitem{hivon} E.~Hivon, K.~M.~Gorski, C.~B.~Netterfield, B.~P.~Crill, S.~Prunet and F.~Hansen, 
Astrophys.\ J.\ {\bf 567}, 2 (2002)
[astro-ph/0105302].


\bibitem{hansen02} F.~K.~Hansen, K.~M.~Gorski and E.~Hivon, 
Mon.\ Not.\ Roy.\ Astron.\ Soc.\ {\bf 336} (2002) 1304
[astro-ph/0207464].



\bibitem{PlanckNG} P.~A.~R.~Ade {\it et al.} [Planck Collaboration], 
arXiv:1303.5084 [astro-ph.CO].



\bibitem{planckcosmo}  P.~A.~R.~Ade {\it et al.} [Planck Collaboration], 
arXiv:1303.5076 [astro-ph.CO].

 



\end{thebibliography}
\end{document}